\documentclass[aps,pre,preprint,superscriptaddress,showkeys,showpacs,amssymb]{revtex4}
\usepackage{graphicx}
\usepackage{graphics}
\usepackage{amsmath}
\usepackage{tabularx}
\usepackage{color}
\usepackage{doi}
\usepackage{textcomp}
\usepackage{dcolumn}
\usepackage{bm}
\newcolumntype{Y}{>{\centering\arraybackslash}X}

\begin{document}

\title{Phase Diagram of the Contact Process on Barabasi-Albert Networks}

\author{D. S. M. Alencar}
\affiliation{Departamento de F\'{\i}sica, Universidade Federal do Piau\'{i}, 57072-970, Teresina - PI, Brazil}
\author{T. F. A. Alves}
\affiliation{Departamento de F\'{\i}sica, Universidade Federal do Piau\'{i}, 57072-970, Teresina - PI, Brazil}
\author{G. A. Alves}
\affiliation{Departamento de F\'{i}sica, Universidade Estadual do Piau\'{i}, 64002-150, Teresina - PI, Brazil}
\author{R. S. Ferreira}
\affiliation{Departamento de Ci\^{e}ncias Exatas e Aplicadas, Universidade Federal de Ouro Preto, 35931-008, Jo\~{a}o Monlevade - MG, Brazil}
\author{A. Macedo-Filho}
\affiliation{Departamento de F\'{i}sica, Universidade Estadual do Piau\'{i}, 64002-150, Teresina - PI, Brazil}
\author{F. W. S. Lima}
\affiliation{Departamento de F\'{\i}sica, Universidade Federal do Piau\'{i}, 57072-970, Teresina - PI, Brazil}

\date{Received: date / Revised version: date}

\begin{abstract}

We show results for the contact process on Barabasi networks. The contact process is a model for an epidemic spreading without permanent immunity that has an absorbing state. For finite lattices, the absorbing state is the true stationary state, which leads to the need for simulation of quasi-stationary states, which we did in two ways: reactivation by inserting spontaneous infected individuals, or by the quasi-stationary method, where we store a list of active states to continue the simulation when the system visits the absorbing state. The system presents an absorbing phase transition where the critical behavior obeys the Mean Field exponents $\beta=1$, $\gamma'=0$, and $\nu=2$. However, the different quasi-stationary states present distinct finite-size logarithmic corrections. We also report the critical thresholds of the model as a linear function of the network connectivity inverse $1/z$, and the extrapolation of the critical threshold function for $z \to \infty$ yields the basic reproduction number $R_0=1$ of the complete graph, as expected. Decreasing the network connectivity leads to the increase of the critical basic reproduction number $R_0$ for this model.

\end{abstract}


\maketitle

\section{Introduction}

Motivated by recent studies on immunity duration against seasonal coronaviruses\cite{Fontanet-2021,Edridge-2021} and the surge of new variants\cite{Callaway-2021}, we consider the contact process\cite{Harris-1974, Ferreira-2016} coupled to the Barabasi-Albert\cite{Cohen-2010, Barabasi-2016} networks. Since we experience the advance of Covid-19 worldwide, we observed that it is spreading out mainly over scale-free networks like airline connections. The most fundamental scale-free model is that reported by Barabasi and Albert\cite{Cohen-2010, Barabasi-2016}. In this work, we revisit this network structure.

Running on top of the network structure, we use the dynamics of the well-known contact process, the simplest epidemic-like model allowing absorbing configurations\cite{Boguna-2013, Pastorsatorras-2015} to describe epidemic spreading without permanent immunity\cite{Dickman-1999, Hinrichsen-2000, Hinrichsen-2006, Odor-2004, Henkel-2008}. The contact process presents an absorbing phase transition, on the directed percolation universality class\cite{Broadbent-1957, Odor-2004, Hinrichsen-2006, Henkel-2008}. In addition, the contact process is a stochastic epidemic process, derived from the master equation by the constant rate Poissonian process approximation\cite{Dorogovtsev-2008, Pastorsatorras-2015}.

Coupling epidemic processes on networks adds some realism at describing an epidemic spreading because networks are ubiquitous in human relationships, represented by bonds between the network nodes\cite{Dorogovtsev-2003, Barabasi-2016}. There are many examples of real scale-free networks, which we can mention the human sexual contacts\cite{Liljeros-2001}, the world wide web\cite{Barabasi-1999, Dorogovtsev-2003, Barabasi-2016}, the transport network\cite{Barrat-2004}, the scientific article citations\cite{Price-1965, Redner-1998}, and the scientific collaborations\cite{Newman-2001, Barabasi-2002}. Possibly, the most known example of a scale-free network model is the Barabasi-Albert (BA) model\cite{Barabasi-1999, Albert-2002, Newman-2002, Dorogovtsev-2003, Barrat-2004, Boccaletti-2006, Barrat-2008, Cohen-2010, Barabasi-2016}. 

We can expect a change in the critical behavior of the contact process when coupled to scale-free networks\cite{Dorogovtsev-2008, Cohen-2010, Odor-2012, Barabasi-2016}. A possible change is a system starting to display non-universal scaling where the critical exponents are functions of the degree distribution exponent\cite{Hong-2007, Wittmann-2014, Ferreira-2011, Lima-2006, Krawiecki-2018, Krawiecki-2018-2, Krawiecki-2019}. In this way, our objective is to characterize the stationary critical behavior of the contact process on Barabasi-Albert networks. Also important is the effect of the network connectivity $z$, defined as the mean number of connections. Decreasing the network connectivity can be interpreted as the individuals adopting a social distancing behavior.

This paper is organized as follows. We describe the model in section \ref{sec:model}. We show details on estimating the critical threshold, and the critical exponents in section \ref{sec:results}. Finally, we present our conclusions in section \ref{sec:conclusions}.

\section{Model and Implementation\label{sec:model}}

In the contact process, a network with $N$ nodes represents a group of $N$ interconnected individuals, each one placed at its respective node. The connections between the individuals are represented by the network edges. The epidemics are spread by contamination of a susceptible individual by one of its infected neighbors, placed in another node connected by an edge.

To grow Barabasi-Albert networks with size $N$ and connectivity $z$, where $z \ll N$, one should start from a complete graph with $z+1$ nodes and add one node at a time where every newly added node $j$ will connect with $z$ randomly chosen previously added nodes in the growing process. Connections are chosen by preferential attachment, i.e., the newly added node $j$ will connect with a previously added node $i$ with a probability $W_i$ proportional to its degree $k_i$. We forbid multiple bonds between the same nodes. The Barabasi-Albert model is a random scale-free model in the sense of unbound fluctuations of the connectivity $z$, which is defined as the average degree\cite{Cohen-2010, Barabasi-2016}. 

We define the contact process as usual\cite{Harris-1974}. The kinetic Monte-Carlo dynamic rules for the synchronous version of the contact process are summarized as follows:
\begin{enumerate}
\item \textbf{Initialization:} The network state is given by $N$ stochastic variables $(\psi_1,\psi_2,...,\psi_N)$ where $\psi_i = 0$ if the node $i$ is susceptible, and $\psi_i = 1$ if the node $i$ is infected. In the time $t=0$, half of the nodes are randomly chosen to be infected. 
We initialize another variable with a null value to count the number of visits $N_{\mathrm{r}}$ to the absorbing state, which is all nodes being susceptibles.
\item \textbf{Evolution step:} One network node is randomly selected and
   \begin{itemize}
      \item If the node is susceptible, we randomly select one of its neighbors and if the selected neighbor is infected, the node becomes infected with a contamination rate $\mu=1$;
      \item However, if the node is infected, it can recover with a recovering rate $\lambda$.
   \end{itemize}
The time lapse of each update is $\Delta t = \alpha/N$. For simplicity, we can choose $\alpha=1$.
\item \textbf{Reactivation of the dynamics:} The contact process has an absorbing state, where all individuals are susceptible. At finite lattices, the only true stationary state is the absorbing one\cite{Dickman-2005}. To prevent the system from visiting the absorbing state, we should reactivate the dynamics, by generating quasi-stationary states. This can be done in two distinct and equivalent ways\cite{Pruessner-2007}:
   \begin{itemize}
      \item \textbf{Reactivation method\cite{Alves-2018, Mota-2018}:} If there is no infected node in the entire network, we increase $N_{\mathrm{r}}$ by one unit, and we randomly infect one node of the network to continue the simulation. Effects of reactivation on the stationary state can be measured by the reactivating field $h_{\mathrm{r}} = \frac{N_{\mathrm{r}}}{Nt}$
which is the average of inserted particles (i.e., spontaneously infected individuals) and scales as $1/N$ in the absorbing phase, vanishing in the thermodynamic limit;
      \item \textbf{Quasi-stationary (QS) method\cite{Oliveira-2005, Dickman-2005}:} We store a list of $N_\text{list}$ active states and update the list at a predefined number of steps $N_\text{steps}$ by replacing one randomly selected state in the list with the actual system state if it is an active one. However, if the system falls in the absorbing state, we randomly select one stored active state in the list to replace the actual absorbing state and continue the dynamics. In order to reproduce simulation results with QS method, one should inform the relevant $N_\text{list}$ and $N_\text{steps}$ parameters. In our simulations, we used $N_\text{steps}=N$ (the list of active states is updated at each time unit), and $N_\text{list} = N$;
   \end{itemize}
\item \textbf{Iteration:} Steps (2) and (3) should be repeated a predefined number of MC steps to let the system evolve to the stationary state. After that, we can continue to iterate steps (2) and (3) to collect the time series of the relevant observables presented in the section \ref{sec:results}.
\end{enumerate}
Note that we used the recovering rate $\lambda$ as the control parameter. However, if one decides to use the contamination rate as the control parameter, they should change the contamination rate to $1/\lambda$, and the recovering rate to $1$ to compare the results with the ones presented here. The basic reproduction number $R_0$ is defined as the ratio of the contamination and recovering rates and is given in terms of the control parameter as $R_0=1/\lambda$.

\section{Results\label{sec:results}}

The main observable is the fraction of infected nodes
\begin{equation}
  \rho = \frac{1}{N}\sum_i^N \psi_i,
\label{infection-concentration}
\end{equation}
that vanishes at the absorbing phase. From the fraction of infected nodes, we can calculate the following averages from the time series of $\rho$ given in Eq.(\ref{infection-concentration})
\begin{eqnarray}
U &=&\left[\frac{\left< \rho^{2} \right>\left< \rho^{3} \right>  - \left< \rho^{ } \right>\left< \rho^{2} \right>^{2}}
          {\left< \rho^{ } \right>\left< \rho^{4} \right>  - \left< \rho^{ } \right>\left< \rho^{2} \right>^{2}} \right],
          \nonumber \\
P &=& \left[\left< \rho \right>\right], \nonumber \\
\Delta &=& N \left[ \left< \rho^{2} \right> - \left< \rho^{ } \right>^{2} \right],
\label{cp-averages}
\end{eqnarray}
where $\left[\cdots\right]$ means a quenched average on the random network realizations, and $\left< \cdots \right>$ means a time series average. $U$ is the $5$-order cumulant for directed percolation\cite{Lubeck-2002, Jansen-2007, Henkel-2008}, $P$ is the order parameter of the active-absorbing transition, and $\Delta$ is the order parameter fluctuation. All averages are given as functions of the recovering rate $\lambda$. The cumulant $U$ should be universal on the critical threshold in the presence of an external field, meaning that the curves for the cumulant should cross on the critical threshold when using the reactivation method. For the QS method, we can estimate the critical threshold by the QS moment ratio, given by\cite{Oliveira-2005}
\begin{equation}
m = \left[\frac{\left< \rho^{2} \right>}{\left< \rho \right>^{2}}\right].
\label{qs-cumulant}
\end{equation}

Close to the critical threshold $\lambda_c$, we conjecture that the $5$-order cumulant, the order parameter, and its fluctuation should obey the following finite-size scaling (FSS) relations\cite{kenna-2012, kenna-2006-1, kenna-2006-2, palchykov-2010}
 \begin{eqnarray}
U &\approx& f_{U}\left[N^{1/\nu}\left( \ln N \right)^{-\widetilde{\lambda}}\left(\lambda-\lambda_{c}\right)\right], \nonumber \\
P &\approx& N^{-\beta/\nu}\left( \ln N \right)^{-\widetilde{\beta}}f_P\left[N^{1/\nu}\left( \ln N \right)^{-\widetilde{\lambda}}\left(\lambda-\lambda_{c}\right)\right], \nonumber \\
\Delta &\approx& N^{\gamma'/\nu} \left( \ln N \right)^{\widetilde{\gamma}'}f_{\Delta}\left[N^{1/\nu}\left( \ln N \right)^{-\widetilde{\lambda}}\left(\lambda-\lambda_{c}\right)\right],
\label{fss-relations}
\end{eqnarray}
where $\beta/\nu = 1/2$, and $\gamma'/\nu = 0$ are the Mean Field critical exponent ratios. The exponent $\nu$ is the shift exponent\cite{Hong-2007, Wittmann-2014} that obeys $\nu = d_c \nu_\perp$ where $\nu_\perp$ is the spatial correlation exponent and $d_c=4$ is the upper critical dimension, with $1/\nu = 1/2$. Expressions (\ref{fss-relations}) account for logarithmic corrections, and our simulation results with reactivation field are compatible with the pseudo-exponents $\widetilde{\lambda}=1/4$, $\widetilde{\beta}=1$, and $\widetilde{\gamma}'=1/8$. Results for the pseudo-exponent corrections can change with the quasi-stationary state, and for the QS method, we conjecture $\widetilde{\lambda}=1/2$, $\widetilde{\beta}=-1/2$, and $\widetilde{\gamma}'=5/4$. QS moment ratio $m$ obeys the same FSS behavior of $U$, however, with a distinct correction pseudo-exponent. 

We simulated the dynamics with the reactivation method on Barabasi-Albert networks with different connectivities $z$ to investigate how the critical thresholds depend on them. We performed a Monte Carlo simulation on networks with sizes: $N=2500$, $N=3600$, $N=4900$, $N=6400$, $N=8100$, and $N=10000$ in order to obtain the averages given in Eq.(\ref{cp-averages}). For each size $N$, and connectivity $z$, we simulated $128$ random network realizations to make quench averages. For each network replica, we considered $10^5$ Markov chain Monte-Carlo (MCMC) steps to let the system evolve to a stationary state and another $10^5$ MCMC steps to collect a time series of $10^5$ values of the fraction of infected nodes. From the time series, we calculated the averages written on Eq.(\ref{cp-averages}), and its respective error bars\cite{Tukey-1958}.

Regarding the CP dynamics with QS method, we considered only the case $z=4$, where we simulated networks with sizes: $N=4900$, $N=6400$, $N=8100$, $N=10000$, $N=12100$, and $N=14400$ in order to obtain the order parameter $P$, and its fluctuations $\Delta$, given in Eq.(\ref{cp-averages}). We also calculated the QS moment ratio $m$ given in Eq.(\ref{qs-cumulant}). We simulated $160$ random network realizations to make quench averages with $2 \cdot 10^5$ MCMC steps, where we discarded the first $10^5$ steps for each network replica.

\begin{figure}[p]
\begin{center}
\includegraphics[scale=0.16]{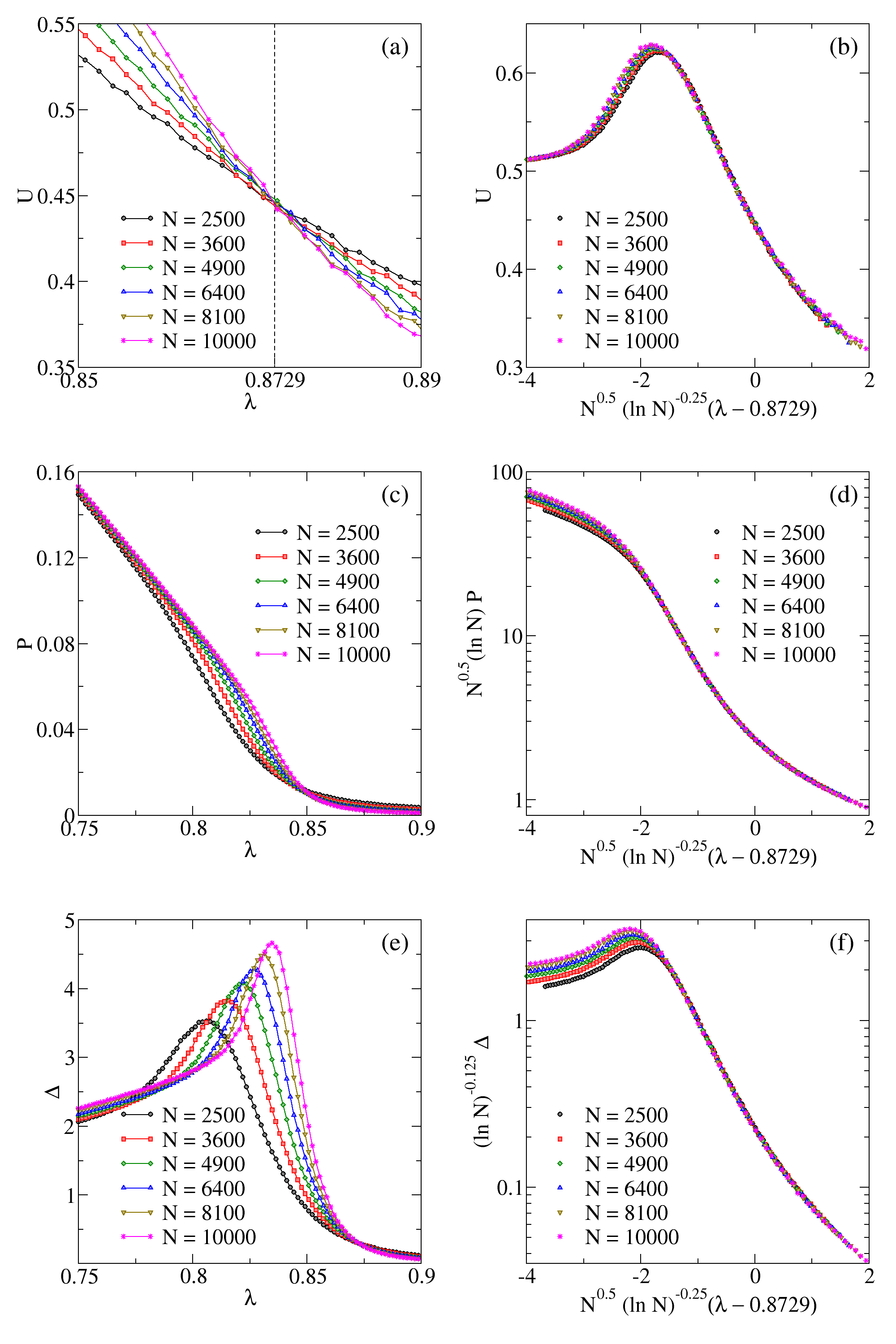} 
\end{center}
\caption{(Color Online) Results for the Contact Process on Barabasi-Albert networks with connectivity $z=4$ of different sizes $N$, where we used the reactivation method. In panels (a), (c) and (e), we show the 5-order cumulant $U$, the infection concentration $P$, and its fluctuation $\Delta$ written in Eq. (\ref{cp-averages}). In panels (b), (d) and (f), we show the scaled plots of $U$, $P$, and $\Delta$ according to the FSS relations written in Eq.(\ref{fss-relations}). The cumulants for different lattice sizes cross on the collective threshold, estimated at $\lambda_c = 0.8729(5)$. The data collapses are compatible with the Mean Field critical exponent ratios and pseudo-correction exponents presented in the section \ref{sec:results}. Statistical errors are smaller than the symbols.}
\label{cp-z=4-results}
\end{figure}

We show results for the Barabasi-Albert networks with connectivity $z=4$ in Fig.(\ref{cp-z=4-results}). The 5-order cumulant should be independent of the system size at the critical point. By inspecting the panel (a), we can obtain the collective critical threshold $\lambda_c = 0.8729(5)$. We did the same for some network connectivities to collect data of critical thresholds summarized in Tab.(\ref{criticalbehaviortable}). The critical thresholds could be refined by inspection of data collapses according to expressions (\ref{fss-relations}) and we obtained a measurement error of $\pm 0.0005$ in all cases. In panel (b) we show the data collapse of the 5-order cumulant and the critical behavior is compatible with the same Mean Field exponents of the complex networks and scale-free networks with degree distribution exponent $\gamma>3$\cite{Hong-2007, Ferreira-2011}, and present logarithmic corrections with the pseudo-exponent $\widetilde{\lambda} = 1/4$.
\begin{table}[h]
\begin{center}
\begin{tabularx}{0.4\textwidth}{YY}
\hline
$z$ & $\lambda_c$ \\ 
\hline
   4  & 0.8729(5) \\
   5  & 0.8985(5) \\
   6  & 0.9160(5) \\
   7  & 0.9282(5) \\
   8  & 0.9372(5) \\
   9  & 0.9445(5) \\
   10 & 0.9502(5) \\
   15 & 0.9671(5) \\
   20 & 0.9755(5) \\
\hline                 
\end{tabularx}
\end{center}
\caption{Summary of critical thresholds $\lambda_c$ on Barabasi-Albert networks for some connectivities $z$. The system will be in the active phase for contamination rates $\mu=1$ and recovering rates $\lambda$ smaller than the critical threshold $\lambda_c$. They were obtained by inspection of the data collapses through finite-size scaling relations presented in Eq.(\ref{fss-relations}).}
\label{criticalbehaviortable}
\end{table}

We present the order parameter in panel (c) of Fig.(\ref{cp-z=4-results}). From the curves, we can identify the active phase for recovering rates smaller than the critical threshold $\lambda_c$ and the absorbing phase on the converse. Note that the reactivation procedure destroys the absorbing phase by introducing tails in the curves of the order parameter and the inflection points separate the active and absorbing phases. The reactivating field $h_r$ in the simulations scales as $1/N$ when going deep in the absorbing phase, in a way that the tails are just a perturbation to the order parameter. In panel (d), we show the respective data collapse of the order parameter, which is compatible with the Mean Field critical exponents and logarithmic corrections with pseudo-exponent $\widetilde{\beta}=1$.

We display the order parameter fluctuation $\Delta$ in panel (e). It presents increasing peaks at the inflection point of the order parameter. However, the critical behavior predicted by the Mean Field exponents should be a finite jump corresponding to $\gamma'=0$. The respective data collapse is shown in panel (f), and the order parameter fluctuation at the critical threshold scale as $(\ln N)^{1/8}$, yielding a pseudo-exponent $\widetilde{\gamma'}=1/8$. We can conclude that the peaks are due to logarithm corrections on the order parameter fluctuations.

\begin{figure}[p]
\begin{center}
\includegraphics[scale=0.16]{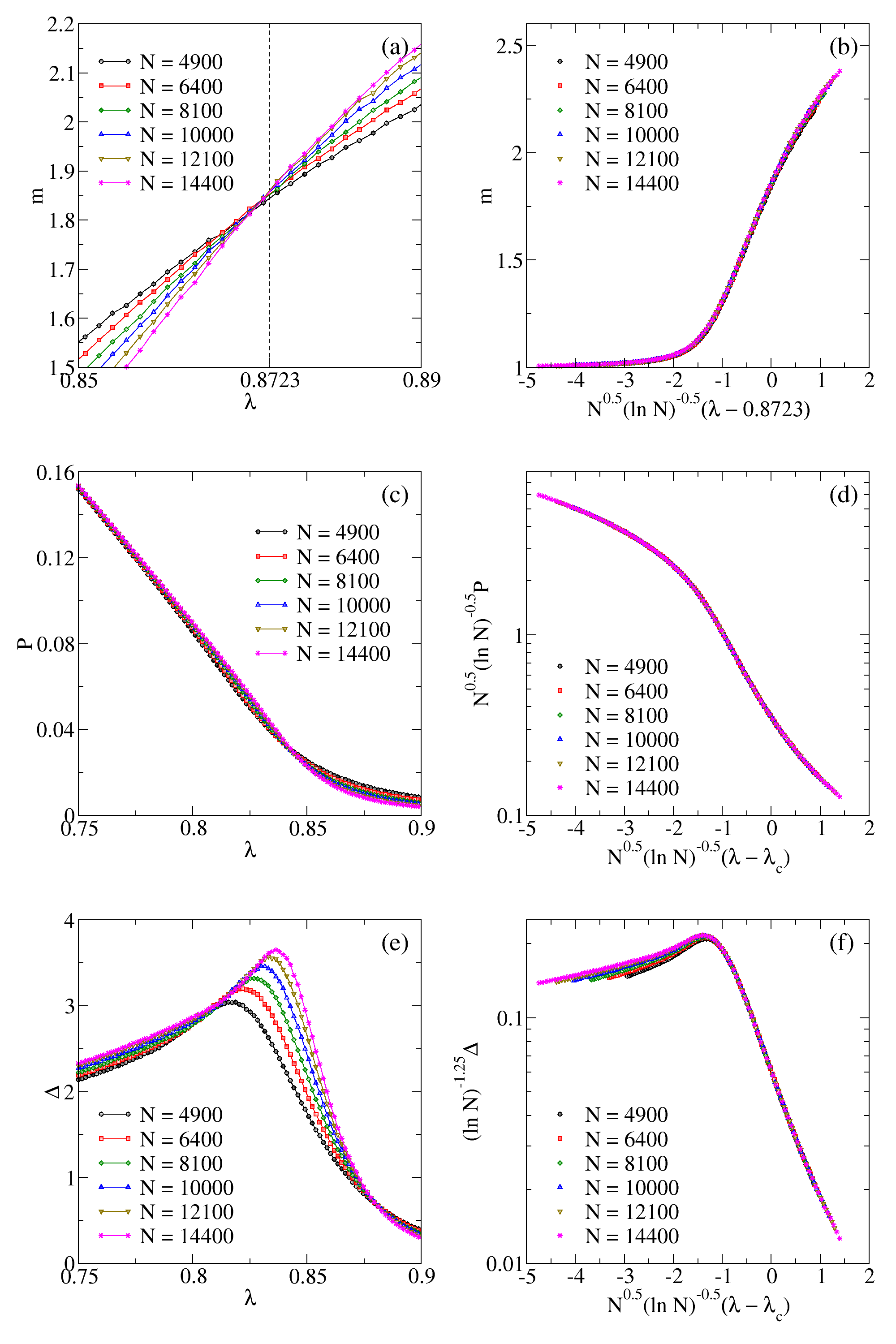} 
\end{center}
\caption{(Color Online) Results for the Contact Process on Barabasi-Albert networks with connectivity $z=4$ of different sizes $N$, where we used the quasi-stationary method. In panels (a), (c) and (e), we show the quasi-stationary moment ratio $m$, the infection concentration $P$, and its fluctuation $\Delta$ written in Eq. (\ref{cp-averages}). In panels (b), (d) and (f), we show the scaled plots of $m$, $P$, and $\Delta$ according to the FSS relations written in Eq.(\ref{fss-relations}). We estimated the critical threshold at $\lambda_c = 0.8723(5)$. The data collapses are compatible with the Mean Field critical exponent ratios and pseudo-correction exponents presented in the section \ref{sec:results}. Error bars are smaller than the symbols.}
\label{cp-z=4-results-QS}
\end{figure}

Now, we compare the effects of distinct QS states by comparing the results obtained with the reactivation procedure with the correspondent results obtained by the QS method, shown in Fig.(\ref{cp-z=4-results-QS}). We estimated the critical threshold at $\lambda_c=0.8723(5)$ where we refined the critical point by inspecting the data collapses. The critical behavior is the same, obeying Mean Field critical exponents. However, the QS state has distinct pseudo-exponent corrections. Data collapses of Fig.(\ref{cp-z=4-results-QS}) are compatible with the values $\widetilde{\lambda}=1/2$, $\widetilde{\beta}=-1/2$, and $\widetilde{\gamma}'=5/4$.

\begin{figure}[h]
\begin{center}
\includegraphics[scale=0.3]{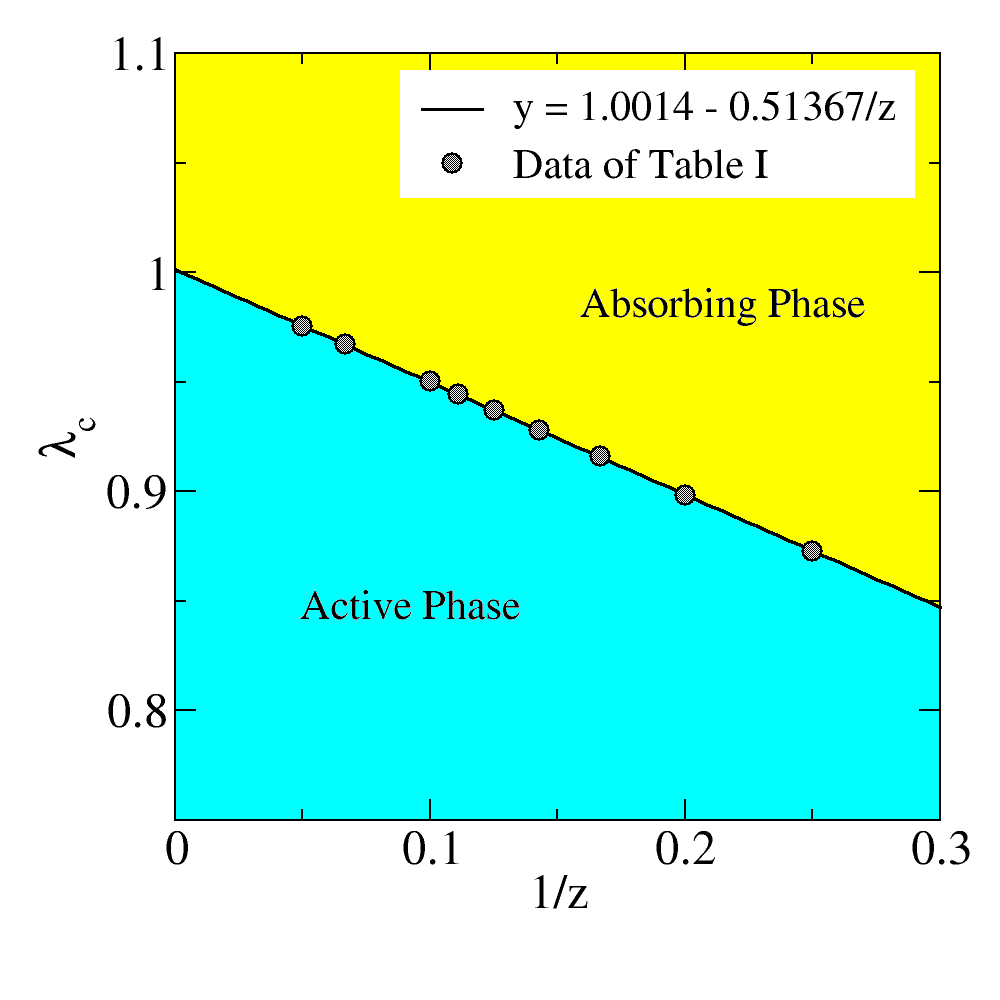} 
\end{center}
\caption{(Color Online) Phase diagram of the contact process on the Barabasi-Albert networks: we show the critical thresholds $\lambda_c$ in Tab.(\ref{criticalbehaviortable}) as functions of the Barabasi-Network connectivity $z$ in gray circles. The critical thresholds are a linear function of $1/z$ where $\lim_{z\to\infty}\lambda_c = 1$, in a way that the limit of the critical threshold is the same as the complete graph. The linear function defines a phase diagram that separates the active and absorbing regions, shown in cyan and yellow, respectively. Error bars are smaller than the symbols.}
\label{phasediagram}
\end{figure}

Finally, we discuss the critical thresholds $\lambda_c$ as functions of the network connectivity $z$ in Fig.(\ref{phasediagram}). A regression of $\lambda_c$ in terms of $1/z$ reveals a straight line that separates the active and absorbing phases. An analogous result was obtained for a kinetic consensus formation model, where the critical thresholds have the same linear dependence on the inverse of the network connectivity\cite{Alves-2020}. Particularly interesting is the limit of the fully connected graph $z\to\infty$ where the basic reproduction number $R_0$ assumes the critical value $R_0=1$ which separates the active phase for $R_0>1$ and the absorbing phase for $R_0\leq 1$. In this way, the control parameter $\lambda = 1/R_0$ has a critical value for the fully connected graph given by $\lim_{z\to\infty}\lambda_c = 1$. Indeed, this is compatible with the extrapolation of the linear regression shown in Fig.(\ref{phasediagram}).

\section{Conclusions\label{sec:conclusions}}

We presented the stationary critical behavior of the contact process on Barabasi-Albert networks. Our data collapses are compatible with Mean Field critical exponents $\beta=1$, $\gamma'=0$ and $\nu=2$ and the existence of logarithmic corrections to the shift scaling when using the reactivation method, i.e., on the $5$-order Binder cumulant, that are proportional to $(\ln N)^{-1/4}$. Logarithmic corrections are also present in the order parameter scaling, proportional to $(\ln N)^{-1}$. Finally, the fluctuation of the order parameter at the critical threshold scales as $(\ln N)^{1/8}$, instead of presenting a finite jump.

By simulating the same dynamics by using quasi-stationary states generated by the QS method, we found the same critical behavior in the Mean Field regime. However, the pseudo-exponents are distinct in this case. We found a correction in the shift scaling proportional to $(\ln N)^{-1/2}$. Also, there is a logarithmic correction in the order parameter at the critical threshold, proportional to $(\ln N)^{1/2}$. In addition, the order parameter fluctuations at the critical threshold scales as $(\ln N)^{5/4}$.

We obtained the phase diagram of the system, i.e., the critical thresholds as functions of the network connectivity. Linear regression of data of critical thresholds reveals a linear function of $1/z$, separating the active and absorbing regions. The extrapolation of the critical threshold in the limit $z\to \infty$ yields the critical basic reproduction number of the complete graph $R_0 = 1/\lambda_c = 1$, as expected. An analogous result was found for a kinetic consensus formation model, where the critical thresholds have the same linear dependence on the inverse of the network connectivity\cite{Alves-2020}. Note that the phase diagram permits us to conclude the social distancing effectiveness in epidemic control. Reducing the connectivity of the network can be interpreted as avoiding social contacts and the direct consequence is to increase the minimal $R_0=1/\lambda_c$ that allows the epidemics to survive.

\section{Acknowledgments}

We would like to thank CAPES (Coordena\c{c}\~{a}o de Aperfei\c{c}oamento de Pessoal de N\'{\i}vel Superior), CNPq (Conselho Nacional de Desenvolvimento Cient\'{\i}fico e tecnol\'{o}gico), FUNCAP (Funda\c{c}\~{a}o Cearense de Apoio ao Desenvolvimento Cient\'{\i}fico e Tecnol\'{o}gico) and FAPEPI (Funda\c{c}\~{a}o de Amparo a Pesquisa do Estado do Piau\'{\i}) for the financial support. We acknowledge the use of Dietrich Stauffer Computational Physics Lab, Teresina, Brazil, and Laborat\'{o}rio de F\'{\i}sica Te\'{o}rica e Modelagem Computacional - LFTMC, Teresina, Brazil, where the numerical simulations were performed.

\bibliography{textv1}

\end{document}